\pdfoutput=1
\documentclass[
    ,final            
  ]
  {aipproc}

\layoutstyle{6x9}

\usepackage{hyperref}

\def\hbabar{\mbox{{\fontsize{20}{11}\bf\sl B}\hspace{-0.14em}{\fontsize{18}{11}\bf\sl A}\hspace{-0.3em}
{\fontsize{20}{11}\bf\sl B}\hspace{-0.14em}{\fontsize{18}{11}\bf\sl
  A\hspace{-0.04em}R}}}

\def\lbabar{\mbox{{\fontsize{12}{11}\sl B}\hspace{-0.35em} {\fontsize{10}{11}\sl A}\hspace{-0.03em
}{\fontsize{12}{11}\sl B}\hspace{-0.35em} {\fontsize{10}{11}\sl A\hspace{-0.02em}R}}}
\def\babar{\mbox{\sl B\hspace{-0.37em} {\small\sl A}\hspace{-0.3em}
    \sl B\hspace{-0.37em} {\small\sl A\hspace{-0.02em}R}}}
\def\sbabar{\mbox{{\fontsize{9}{11}\sl B\hspace{-0.4em}
      \fontsize{7pt}{11pt}\sl A}\hspace{-0.35em} \fontsize{9}{11}\sl B\hspace{-0.4em} {\fontsize{7pt}{11pt}\sl
      A\hspace{-0.02em}R}}}

\def\sss{\scriptscriptstyle}
\def\barpd{{\raise.35ex\hbox{${ \sss (}$}}--{\raise.35ex\hbox{${\sss )}$}}}
\def\kstar{\hbox{$K^{*0}$\kern-1.5em\raise1.5ex\hbox{\barpd}}\:}

\begin{document}

\title{Lepton-flavor-violating $\tau$ decays at \hbabar}

\classification{13.35.Dx, 14.60.Fg, 11.30.Hv}
\keywords      {lepton-flavor violation, $\tau$ decays}

\author{Giovanni Marchiori, representing the \lbabar\ Collaboration}{
  address={Laboratoire de Physique Nucl\'eaire et de Hautes Energies,
    IN2P3/CNRS, F-75252 Paris, France },
  altaddress={giovanni.marchiori@lpnhe.in2p3.fr}
}

\begin{abstract}
We present the most recent searches for lepton-flavor-violating (LFV)
$\tau$ decays in \sbabar.
We find no evidence of $\tau$ decaying to three charged leptons or to
a charged lepton and a neutral meson ($K^0_S$, $\rho$, $\phi$,
$K^{*0}$, $\overline{K}$$^{*0}$), and set upper limits on the 
corresponding branching fractions (BF) between 1.8 and 19
$\times 10^{-8}$ at 90\% confidence level (CL).
\end{abstract}

\maketitle


\section{Theoretical relevance}
The experimental observation of LFV $\tau$ and $\mu$ decays
would provide unambiguous evidence of New Physics. 
In the Standard Model (SM) in fact they proceed through diagrams
with neutrinos in the loops: the strong GIM suppression, due
to the small neutrino masses, leads to branching fractions 
that are below any achievable experimental sensitivity. 
On the other hand, in supersymmetric (SUSY) models, LFV $\tau$ and
$\mu$ decays can receive significant contributions from diagrams
containing SUSY particles in the loops, via slepton mixing. 
Depending on the model and the values of the free parameters of the
theory (mass spectra and couplings), LFV $\tau$ decays may have
branching fractions as high as $10^{-7}$ (see e.g. \cite{theo}), thus
within the reach of the \babar\ experiment.

\section{The BaBar data sample}
\babar~\cite{babar:det} is a multi-purpose detector operating at the
PEP-II $e^+e^-$ collider at SLAC.
Charged particles' momenta and impact parameters are measured by a
tracking system, consisting of a silicon strip detector and a gaseous
drift chamber in a solenoidal magnetic field of 1.5T.
A Tl-doped CsI calorimeter identifies photons and
electrons and determines their energy.
Muons are identified by resistive plate
chambers and limited streamer tubes installed in the gaps of the iron
that contains the solenoid flux return.
Kaon/pion discrimination is based on the opening angle and photon
yield of the Cherenkov light emitted in synthetic quartz bars, and 
on ionization energy loss in the tracking devices.
Through the years 2000-2008 \babar\ has collected almost 1 billion
$\tau$ leptons, produced in $e^+e^-{\to}\tau^+\tau^-$ events.
Here we consider only data taken at $\sqrt{s}\approx 10.58$ GeV (about
90\% of the total $\tau$ sample).
At this center-of-mass (CM) energy the $\tau\tau$ cross section,
$\approx 0.92$ nb, is comparable to that of dimuon production and
about 1/5 of the hadronic cross section ($e^+e^-{\to}q\bar{q}$,
$q=u,d,s,c,b$). The effective cross section for Bhabha events where at
least one $e^\pm$ is within the acceptance of the calorimeter is about
50 times larger.

\section{Selection criteria}
We search for the LFV decays $\tau \to ll'l''$,
$lV^0$ and $lK^0_S$, where $l^($$'$$^,$$''^)=(e,\mu)$,
$V=(\phi,\rho,K^*,\overline{K}^*)$, 
$\phi{\to}K^+K^-$, $\rho^0{\to}\pi^+\pi^-$,
$K^{*0}{\to}K^+\pi^-$, $\overline{K}$$^{*0}{\to}K^-\pi^+$,
$K^0_S{\to}\pi^+\pi^-$. 
We look for $\tau$-pair events where one ({\em signal}) $\tau$ decays
to a fully reconstructed LFV final state and the
other ({\em tag}) $\tau$ decays to a partially reconstructed SM final state.
We consider only one-prong tag $\tau$ decays (BF $\approx 85\%$),
except for the $eK^0_S$ analysis, where we include also 
three-prong decays (BF $\approx 15\%$).
We thus divide the event in two hemispheres in the CM frame, by
means of a plane perpendicular to the event's thrust axis, and require that
the 3 charged particles' tracks from the signal $\tau$ and the tracks
(1 or 3) from the tag $\tau$ decay belong to different hemispheres. 
All tracks must be well reconstructed, within the fiducial
volume of the detector, and should have zero total charge.
Pairs of tracks consistent with a photon conversion are removed in
order to suppress radiative QED backgrounds.
The invariant masses of the $K^0_S$ and $V^0$ daughters are required
to be close to the mass of the originating mother;
the $K^0_S$ decay vertex should be displaced from the interaction point.
Tight $e/\mu/\pi/K$ identification criteria, with $\approx 1\%$ 
misidentification probability, are applied to the signal $\tau$
daughters. $e/\mu$ vetoes are applied in some cases to the track
from 1-prong tag $\tau$ decays in order to reduce $\mu\mu$ and Bhabha 
backgrounds.
Finally, since the SM tag $\tau$ decay contains undetected
neutrino(s), missing momentum must be different from zero and point 
inside the detector acceptance.

From the measured four-momenta of the three decay
products of the signal $\tau$ we determine its initial four-momentum
and compute the difference $\Delta E \equiv
E^*_{\tau}-\sqrt{s}/2$ between the $\tau$ and the beam energy in the CM, 
and the difference $\Delta M \equiv M^{\rm ec}_{\tau} - M^{\rm WA}_{\tau}$
between the energy-constrained invariant mass of the $\tau$
(after applying the kinematic constraint $\Delta E = 0$)
and the $\tau$ mass world average.
In the $(\Delta M, \Delta E)$ plane, signal should peak around
the origin (with tails at negative $\Delta E$ and positive $\Delta M$
values due to radiation emitted in the initial and final state,
respectively), while backgrounds should be more uniformly distributed.
For each decay mode we define, in the $(\Delta M, \Delta E)$ plane, a
signal region (SR) around the origin, and an external sideband (SB)
region. After applying tight selection requirements in order to reduce
the backgrounds as much as possible, we compare the number of
candidates observed in the SR and the expected background yield,
extrapolated from the number of candidates in the SB.
The SR extent and the selection criteria are optimized, using 
large samples of simulated signal and background
events in addition to background-enriched data control samples, in
order to minimize the expected branching fraction upper limits (UL) in
the background-only hypothesis.

\section{Background extrapolation}
Dominant background sources common to all final states
are misreconstructed SM $\tau\tau$ events and random combination of
tracks in $q\bar{q}$, $q={u,d,s}$ events.
Important background sources are also: Bhabha or dimuon events for
the $ll'l''$ search and $c\bar{c}$ events (in particular $D{\to}V^0l\nu,
K^0_Sl\nu$ decays) for the $lV^0$, $lK^0_S$ searches. 
Simulated events and background-enriched data control
samples (obtained by relaxing or reverting some of the selection criteria)
are used to extract the shapes of the probability density
functions (PDFs) of the various background sources in the $(\Delta M,
  \Delta E)$ plane and their relative abundance. 
The normalization of the total background PDF is fitted to the data in
the SB, and the expected background in the SR is estimated from
integration of the PDF in that range. 
After applying the tight selection criteria, only very few background
events are expected in the SR.
Several kinematic distributions and yields in slices of the SB are
compared between data and simulated events in order to 
evalutate the reliability of the simulation.

\section{Results}
\begin{figure}
  \includegraphics[width=.4\textwidth, height=.16\textheight]{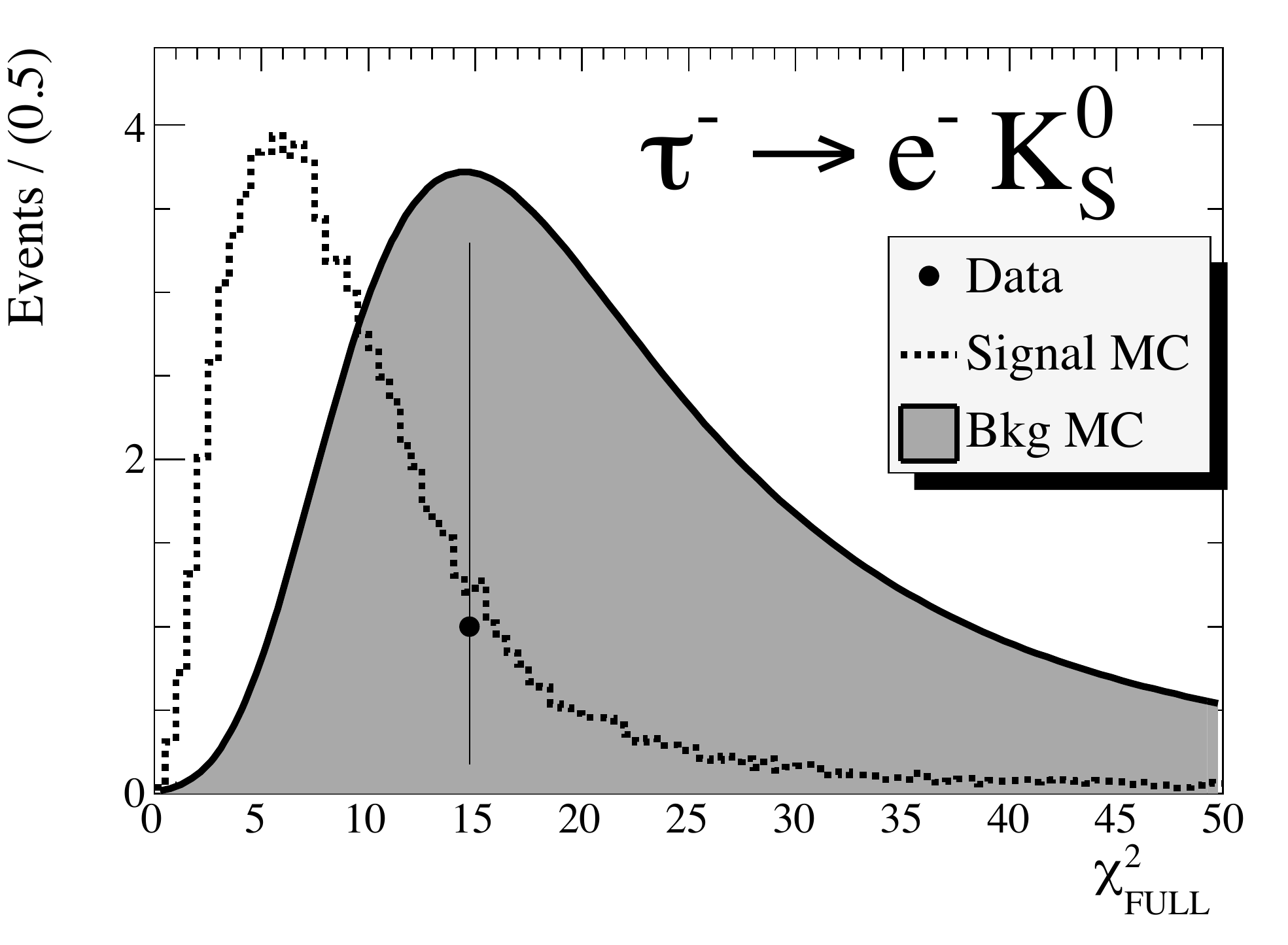}
  \includegraphics[width=.4\textwidth, height=.16\textheight]{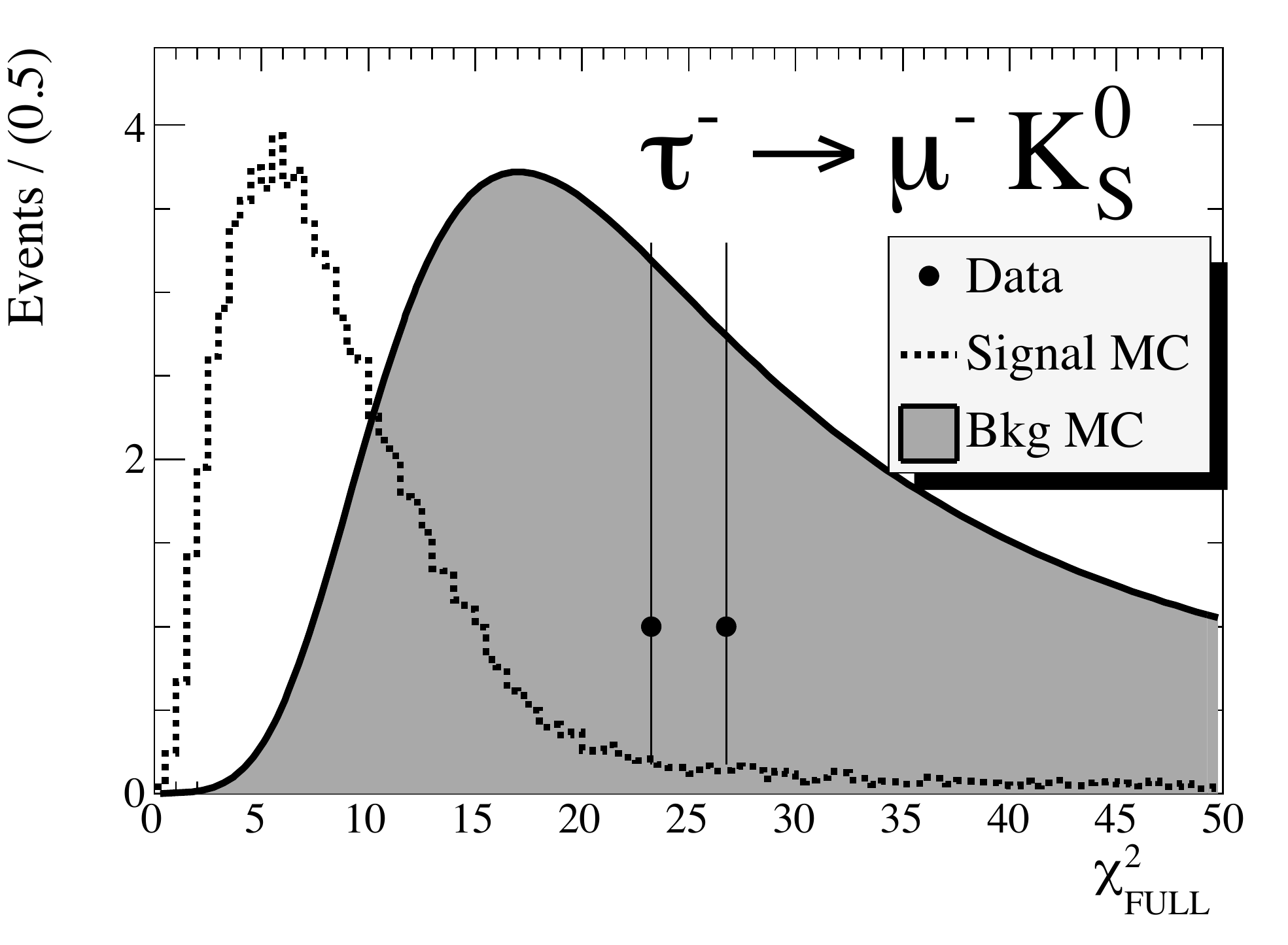}
  \caption{Distribution of the $\chi^2$ of the decay tree fit
    for $\tau\to lK^0_S$ candidates selected in data (dots).
    The dashed line and the shaded histogram represent the expected
    signal and background distributions, respectively (arbitrary
    normalization).}
  \label{fig:results_lKS}
\end{figure}

\begin{figure}
  \centering
  \begin{tabular}{c}
  \includegraphics[width=.6\textwidth, height=.22\textheight]{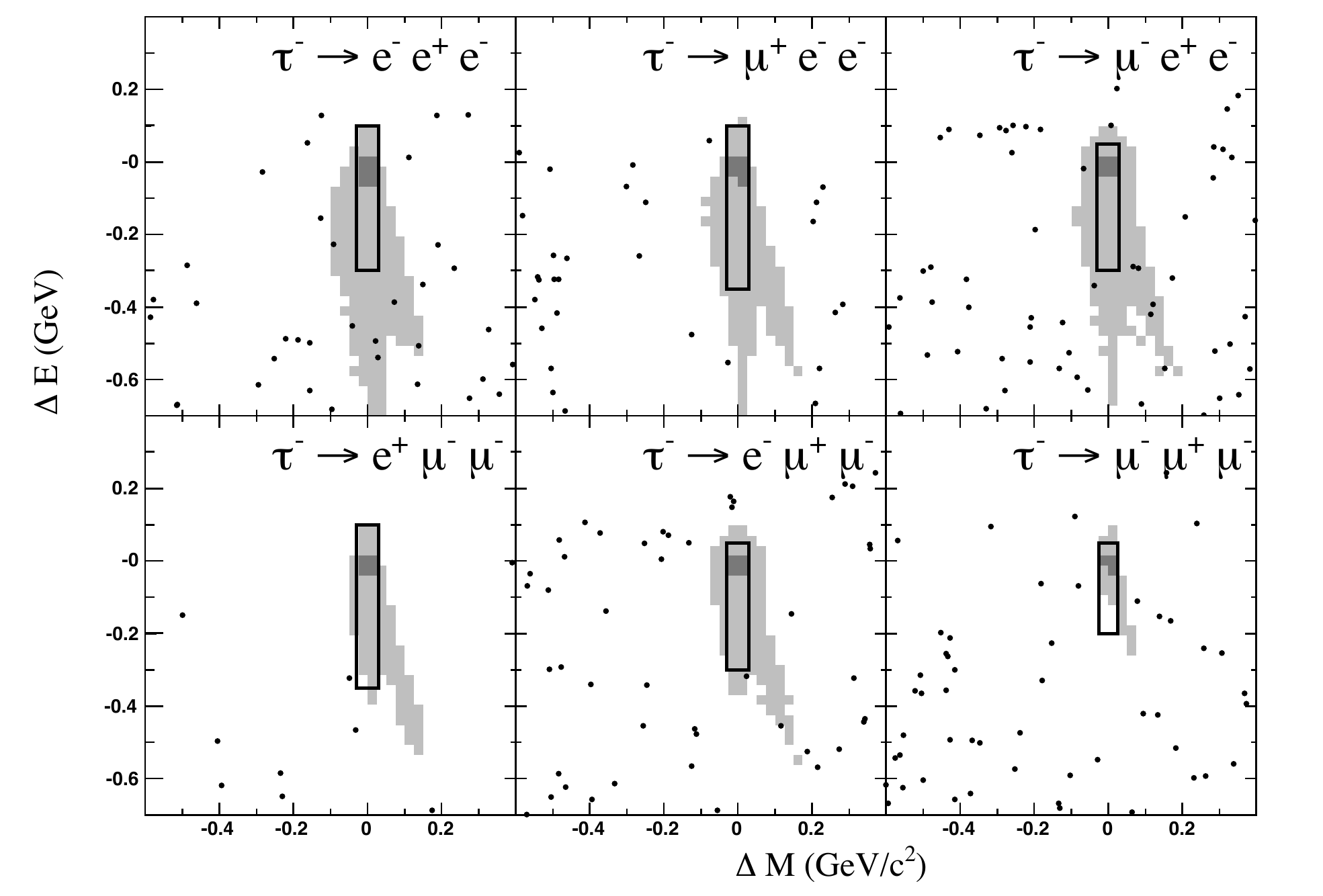} \\
  \includegraphics[width=.8\textwidth, height=.21\textheight]{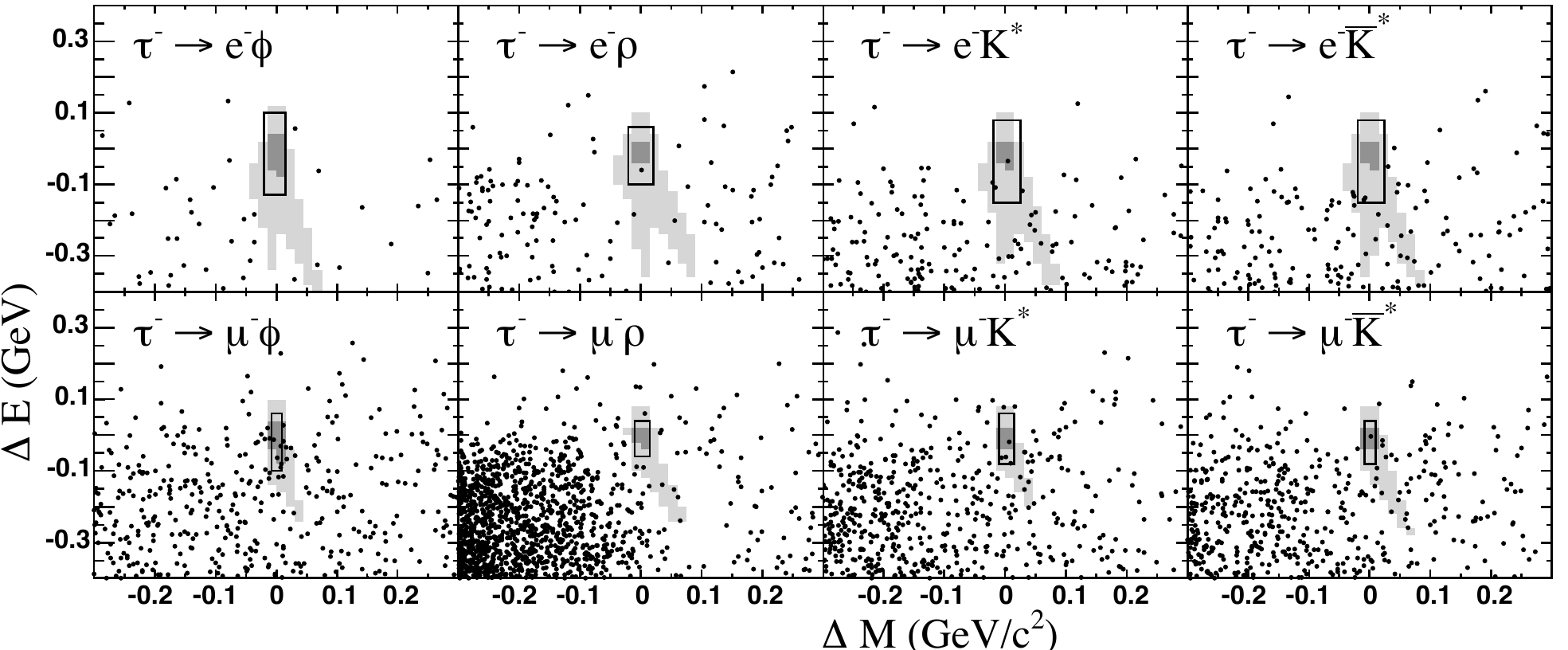}
\end{tabular}
  \caption{$(\Delta M, \Delta E)$ distribution of $\tau\to ll'l''$
      (top) and $\tau\to lV^0$ (bottom) candidates selected in data
      (dots).
      The solid line shows the boundaries of the signal region.
      The dark and light shading indicates contours containing 50\%
      and 90\% of the selected MC signal events, respectively.}
  \label{fig:results_lll_and_lV0}
\end{figure}

For each LFV $\tau$ decay,
the observed number of candidates ($N_{\rm obs}$) agrees with
the expected SM background ($N^{\rm bkg}_{\rm exp}$).
Comparing these quantities, taking into account the number $N_\tau$ of
$\tau$ leptons in the initial sample and the signal selection
efficiency $\varepsilon$ (from simulated events),
we set upper limits between 1.8 and 19 $\times 10^{-8}$ at 90\%
CL on the corresponding branching fractions. We use
either a fully ($ll'l'', lV^0$~\cite{freqUL}) or
``modified'' ($lK^0_S$~\cite{CLs}) frequentist procedure: in the
latter case we compare also the observed and expected (in the
background-only and background+signal hypotheses) distributions of the
$\chi^2$ of the 
geometric and kinematic fit to the whole signal $\tau$ decay tree.
The quoted limits include the systematic uncertainties on $N_\tau$
($\approx1\%$, from the 
luminosity uncertainty), $\varepsilon$ (2-9\%, dominated
 by the uncertainty on the particle-identification efficiency) 
and $N^{\rm bkg}_{\rm exp}$
(typically $<0.3$, from the uncertainties in the background PDF shapes
and the overall background normalization).
The results are summarized in Table~\ref{tab:results} and
illustrated in Figures~\ref{fig:results_lKS} and
\ref{fig:results_lll_and_lV0}. In several cases we improve
previous experimental bounds. 
Some regions of the parameters' space of some SUSY models (like
~\cite{theo}) are excluded by these results.

\begin{table}
\begin{tabular}{lcrrcccc}
\hline
  \tablehead{1}{l}{b}{Decay} &
  \tablehead{1}{c}{b}{$\mathbf N_\tau$} &
  \tablehead{1}{c}{b}{$\mathbf{\varepsilon}$} &
  \tablehead{1}{c}{b}{$\mathbf{N_{exp}^{bkg}}$} &
  \tablehead{1}{c}{b}{$\mathbf{BF_{exp}^{90}}$} &
  \tablehead{1}{c}{b}{$\mathbf{N_{\rm obs}}$} &
  \tablehead{1}{c}{b}{$\mathbf{BF^{90}_{\rm obs}}$} &
  \tablehead{1}{c}{b}{Ref.} \\
   & 
  \tablehead{1}{c}{b}{$(10^6)$} &
  \tablehead{1}{c}{b}{$(\%)$} &  & $(10^{-8})$ &   & $(10^{-8})$ & \\
\hline
$e^\pm e^\pm e^\mp$       & 860 & $ 8.6\phantom{0}\pm 0.2\phantom{0}$ & $0.12\pm 0.02$ & 3.4 & 0 & 2.9 & \cite{babar:lll}  \\
$\mu^\pm e^\pm e^\mp$     & 860 & $ 8.8\phantom{0}\pm 0.5\phantom{0}$ & $0.64\pm 0.19$ & 3.7 & 0 & 2.2 & \cite{babar:lll}  \\
$\mu^\pm e^\mp e^\mp$     & 860 & $12.7\phantom{0}\pm 0.7\phantom{0}$ & $0.34\pm 0.12$ & 2.2 & 0 & 1.8 & \cite{babar:lll}  \\
$e^\pm \mu^\pm \mu^\mp$   & 860 & $ 6.4\phantom{0}\pm 0.4\phantom{0}$ & $0.54\pm 0.14$ & 4.6 & 0 & 3.2 & \cite{babar:lll}  \\
$e^\pm \mu^\mp \mu^\mp$   & 860 & $10.2\phantom{0}\pm 0.6\phantom{0}$ & $0.03\pm 0.02$ & 2.8 & 0 & 2.6 & \cite{babar:lll}  \\
$\mu^\pm \mu^\pm \mu^\mp$ & 860 & $ 6.6\phantom{0}\pm 0.6\phantom{0}$ & $0.44\pm 0.17$ & 4.0 & 0 & 3.3 & \cite{babar:lll}  \\
$e^\pm \phi$             & 830 & $ 6.43\pm 0.16$ & $0.68\pm 0.12$ & 5.0 & 0 & 3.1 & \cite{babar:lV0}  \\
$\mu^\pm \phi$           & 830 & $ 5.18\pm 0.27$ & $2.76\pm 0.16$ & 8.2 & 6 & 19  & \cite{babar:lV0}  \\
$e^\pm \rho^0$           & 830 & $ 7.31\pm 0.18$ & $1.32\pm 0.17$ & 4.9 & 1 & 4.6 & \cite{babar:lV0}  \\
$\mu^\pm \rho^0$         & 830 & $ 4.52\pm 0.41$ & $2.04\pm 0.19$ & 8.9 & 0 & 2.6 & \cite{babar:lV0}  \\
$e^\pm K^{*0}$            & 830 & $ 8.00\pm 0.19$ & $1.65\pm 0.23$ & 4.8 & 2 & 5.9 & \cite{babar:lV0}  \\
$\mu^\pm K^{*0}$          & 830 & $ 4.57\pm 0.36$ & $1.79\pm 0.21$ & 8.5 & 4 & 17 & \cite{babar:lV0}  \\
$e^\pm \overline K^{*0}$  & 830 & $ 7.76\pm 0.18$ & $2.76\pm 0.28$ & 5.4 & 2 & 4.6 & \cite{babar:lV0}  \\
$\mu^\pm \overline K^{*0}$& 830 & $ 4.11\pm 0.32$ & $1.72\pm 0.17$ & 9.3 & 1 & 7.3 & \cite{babar:lV0}  \\
$e^\pm K^0_S$            & 862 & $ 9.4\phantom{0}\pm 0.2\phantom{0}$ & $1.0\phantom{0}\pm 0.4\phantom{0}$ & 3.0 & 1 & 3.3 & \cite{babar:lKS}  \\
$\mu^\pm K^0_S$          & 862 & $ 7.0\phantom{0}\pm 0.4\phantom{0}$ & $5.3\phantom{0}\pm 2.2\phantom{0}$ & 4.8 & 2 & 4.0 & \cite{babar:lKS}  \\
\hline
\end{tabular}
\caption{Number of $\tau$ decays (N$_\tau$), signal efficiency
  ($\varepsilon$), expected background candidates (N$^{\rm
    bkg}_{\rm exp}$) and BF upper limit at 90\% CL (BF$^{\rm
    90}_{\rm exp}$), observed candidates (N$_{\rm obs}$) and 
   branching fraction upper limits (BF$^{\rm
    90}_{\rm obs}$) for each LFV decay mode.}
\label{tab:results}
\end{table}

\bibliographystyle{aipproc}   

\end{document}